\documentclass{eas}
\usepackage{graphicx}
\begin{document}
\title{\large A Thickness of Stellar Disks of Edge-on Galaxies  and
Position of Their Truncation Radii} 
\runningtitle{A.V.Zasov, D.V.Bizyaev: A Thickness of stellar disks \dots}
\author{A.V.Zasov}\address{Sternberg Astronomical Institute, University prosp., 13, 119899, Moscow, Russia}
\author{D.V.Bizyaev}\address{Sternberg Astronomical Institute, University prosp., 13, 119899, Moscow, Russia; Physics Dept., UTEP, El Paso, TX, 79968, USA }

\begin{abstract}
The relationship between the geometrical properties of stellar
disks (a flatness and truncation radius) and the disk
kinematics are considered for edge-on galaxies. It is shown that
the observed thickness of the disks and the approximate constancy
of their thickness along the radius well agrees with the
condition of their marginal local gravitational stability. As a
consequence, those galaxies whose disks are thinner should harbor
more massive dark haloes. The correlation between the de-projeced
central brightness of the disks and their flatness is found (the
low surface brightness disks tend to be the thinniest ones). We
also show that positions of observed photometrically
determined truncation radii $R_{cut}$ for the stellar disks
support the idea of marginal local gravitational stability of
gaseous protodisks at $R =R_{cut}$, and hence the steepening of
photometric profiles may be a result of too inefficient star
formation beyond $R_{cut}$.

\end{abstract}
\maketitle
\section{Introduction}

Stellar disks in spiral  galaxies consist mainly of old stars and
usually include the main part of their stellar content. Being
collisionless, they nevertheless may experience the evolution of
mass distribution - both along the radial and vertical coordinates
due to internal dynamical processes and outer
interactions with neighbour systems. What determines the relative
disk thickness and disk radial extension, remains an open
question. Are these parameters a result of a long lasting
evolution or a product of the very early period of the stellar
disks formation? The analysis of connection between these
parameters and the kinematic properties of galaxies helps to find
an answer.

\section{A thickness of stellar disks of edge-on galaxies}

A flatness of stellar disks differ strongly from one galaxy to
another. As it was found by some authors,  photometric axes
ratioes for flat edge-on galaxies correlate with their
morphological type: disks of late-type galaxies (Sc--Sd) are on
average "thinner" than that of early-type objects (Karachentsev
\etal \cite{kar97}; de Grijs \cite{degr98}; Ma \etal \cite{ma97},
\cite{ma99}).  At the same time the data of the Revised Flat
Galaxy Catalog (Karachentsev \etal, \cite{kar99}, RFGC hereafter)
does not reveal any correlation of axes ratioes with the rotation
velocity or the luminosity. To illustrate it, in Fig.~1 we
compare the $B$-band axes ratio $a/b$ according to the catalog
RFGC with the known HI line width ($W_{50}$) and with the absolute
magnitude $M_B$. Both these  parameters were adopted from the
LEDA catalog. The relationship between the thickness of the disks
and their kinematics, if exists, has to be more sophisticated.

\begin{figure}
\centerline{ \includegraphics[width=13cm,angle=0]{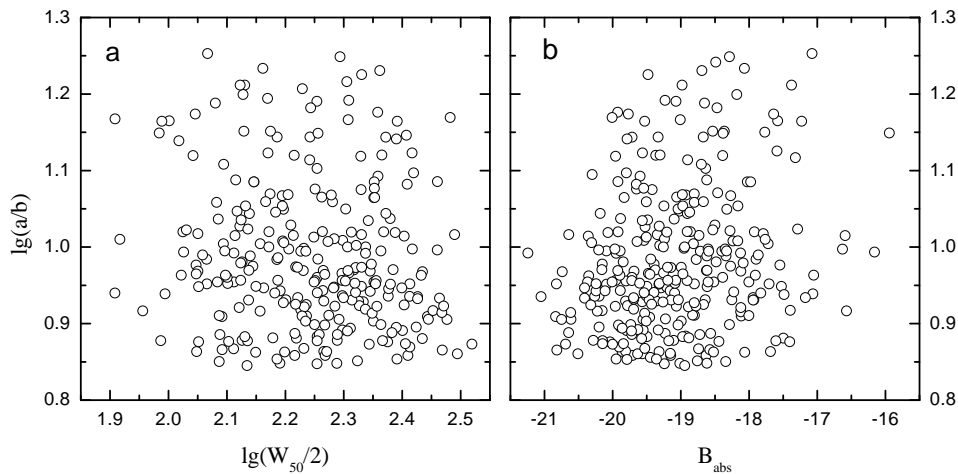} }
\caption{The major to minor axis ratio $a/b$ in the $B$-band
(according to the RFGC catalog) shows no correlation
with the known HI line width ($W_{50}$) and with the absolute
magnitude $M_B$. The figure is adopted from
Zasov \etal (\cite{zasov02}).}
\end{figure}

A thickness of a self-gravitating disk depends on its density and
the stellar vertical velocity dispersion. It is worth noting
that the dynamical heating of a collisionless stellar disk is the
one-way process -- after the increasing of the stellar velocity
dispersion the disk never becomes cooler or thinner again. There
are several heating mechanisms known which might increase the
vertical scale of the disk, such as a scattering of stars either
by massive gas clouds or by other density irregularities,  a
merging of small satellites, or a gravitational interaction with
neighbours. One may easily find the discussion of these
mechanisms in the literature. However although the efficiency of
any of them should depend on the distance from the galactic
center, the observed thickness of disks of edge-on galaxies is
usually nearly constant within a wide interval of radial
distances (although some exceptions may also exist).

Let's imagine for instance that we have a disk where the initial
velocity dispersion of stars in z-direction $C_z$ is very low,
say 3 km/s instead of more realistic value 30 km/s. Let also we
have neither gas clouds nor satellites or close neighbours. Will
a disk stay razor-thin in this case? The answer is definitely NO,
and numerical experiments demonstrate it very clearly: the disk
always heats up within a short time interval (a couple of
rotation periods) as a result of the bending instability which
inevitably develops in a thin disk. The thickness ceases to grow
when some critical value of $C_z$ is reached. This instability
determines the minimum possible thickness for collisionless
disks. The other heating processes, if involved, may only
increase the vertical scalelength of the disk.

As it is known, the bending instability is in some sense an
addition to the classical Jeans instability in the disk plane.
Indeed, the increasing of radial velocity dispersion $C_r$ makes
the disk more stable to Jeans perturbations and less stable to
the bending ones. Hence a growth of $C_r$ leads to the parallel
increasing of the vertical dispersion $C_z$. As a result, one can
expect the proportionality between $C_r$ and $C_z$ in the final
stable state (although, as N-body simulations show, their ratio
may depends on the relative masses of spherical and flat
components, see Mikhailova et al., 2001). The simplified model of
the disk leads to the ratio $C_z/C_r \approx 0.4$ (Polyachenko \&
Shukhman \cite{pol97}), whereas the observations of real galaxies
inferred the values 0.5 - 0.8 with rather large uncertainties. On
the other hand,  the minimal value of $C_r$ is determined by the
local gravitational stability of the disk. In general case the
critical value of $C_r$ is proportional to the surface density
over the epicyclic frequency ratio $\Sigma_{disk} / \kappa$. The
latter in turn is a function of circular velocity $V(R)$, which
makes the minimum possible thickness of the disk connected with
the kinematic properties of a galaxy.

\begin{figure}
\centerline{ \includegraphics[height=12cm,angle=0]{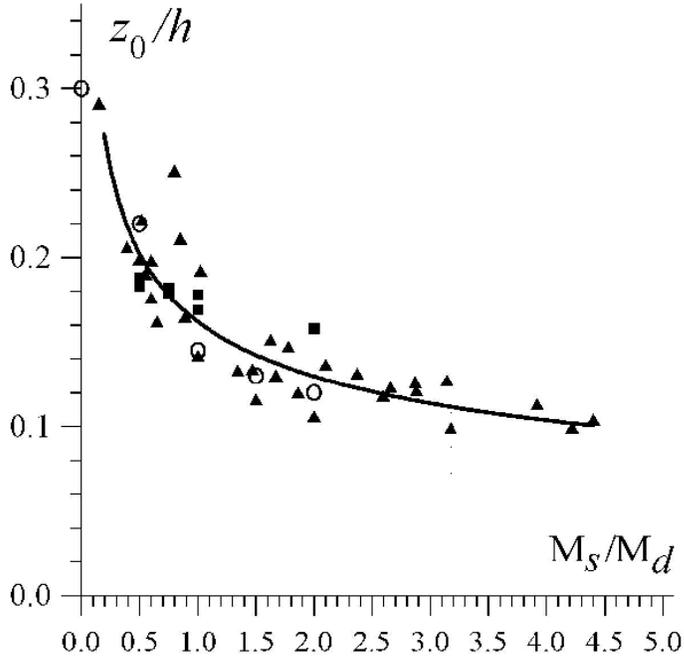} }
\caption{Relation between the ratio of vertical to radial disk
scales and the halo-to-disk mass ratio $M_s/M_d$ obtained from
numerical simulations for the galaxies with the marginally stable
stellar disks. The figure is adopted from Mikhailova \etal~
\cite{mikh01} (their Fig.2). Different symbols correspond to the
models which were constructed for galaxies of different masses and
shapes of their rotation curves.}
\end{figure}

Let us use the simple proportions written for some fixed
dimensionless radius $R/h$ of a galaxy where $h$ is its radial
scalelength:

-- the vertical velocity dispersion $C_z \sim C_r$;

-- the vertical scalelength $Z_0 \sim C_z^2 /\Sigma_{disk}$;

-- the minimal radial velocity dispersion
$C_r \sim \Sigma_{disk}/\kappa$;

-- the total disk mass of a galaxy $M_d \sim \Sigma_{disk} h^2$;

-- the total disk mass of a galaxy $M_t \sim V^2h$, and

-- the epicyclic frequency $\kappa \sim V/h$.

Combining these expressions, one can find that the minimal
relative thickness $Z_0/h$ is expected to be proportional to the
relative mass of the disk: $Z_0/h~\sim~ M_d/M_t$. This conclusion
agrees well with the results of the 3D $N$-body numerical
simulations of collisionless initially thin disks embedded into a
rigid halo (see Zasov \& Morozov \cite{zasov85}, Zasov \etal~
\cite{zasov91}, Mikhailova \etal~ \cite{mikh01}, Zasov \etal~
\cite{zasov02}). Fig.~2 shows the diagram taken from Mikhailova
\etal~ (\cite{mikh01}), where the mass ratio of spherical and
disk components $M_s/M_d$ is compared with the
vertical over radial scalelengths ratio. In accordance with the
expectations, the model galaxies with the most massive halo own
the thinniest stable disks. The question is whether the real
galaxies obey this rule as well.

To verify it, we composed and analysed two samples of edge- on
galaxies with available rotation curves and a surface photometry
(see Zasov \etal~\cite{zasov02} for details). All galaxies were
taken from RFGC and thought to be disk-dominated objects with a
small, if any, bulge contribution to the integrated luminosity.
The nearby galaxies, the Virgo cluster members, and galaxies with
asymmetric or distorted isophotes were excluded from the
consideration.

The first sample includes the objects for which the R-band
surface CCD photometry was performed with the 6m telesope at the
Special Astrophysical Observatory of RAS (Karachentsev \etal~
\cite{kar92}). We estimated the radial scale length $h$ by
fitting the photometric major-axis profiles to the analytical
expression obtained for a model edge-on exponential disk (van der
Kruit\&Searle \cite{vdc81a}, \cite{vdc81b}). Once the radial
scalelength is known, the vertical scale $Z_0$ can be found from
the observed axes ratio. We will call this sample as the
"BTA-sample" hereafter. The second ("2MASS") sample consists of
galaxies whose infrared images had been taken from the 2MASS
survey in the $K_s$ band. The radial and the vertical scales for
the galaxies were restored from their photometric cuts made with
the 2MASS $K_s$-band images (avoiding major axes of the disks to
reduce the influence of the dust) by Bizyaev \& Mitronova
\cite{biz02} (all references may be found in this paper). The
comparison of two samples has revealed that the galaxies in the
$K_s$ -band look systematically thinner than those in the bluer
one (in accordance with the results obtained by de Grjis,
\cite{degr98}), and hence, the samples should be considered
separately.

The starting point for our comparison of the disk thickness with
the relative mass of halo in galaxies is that the presence of a
dark halo always increases the total  mass-to-luminosity ratio
$M/L$ of a galaxy. For definiteness we will estimate all masses
within four disk scalelengths $4h$, which encompasses practically
all stellar mass and luminosity of the galaxy. In the first
approximation the total mass of the galaxy is $M(4h)~=~W_H^2h/G$. The
luminosity was estimated from the photometric parameters of the
disks found after the data processing. The diagrams $M/L$ (that is
$W_H^2h/GL$) versus $h/Z_0$ are shown in Fig.~ 3a,b.

\begin{figure}
\centerline{ \includegraphics[width=13cm]{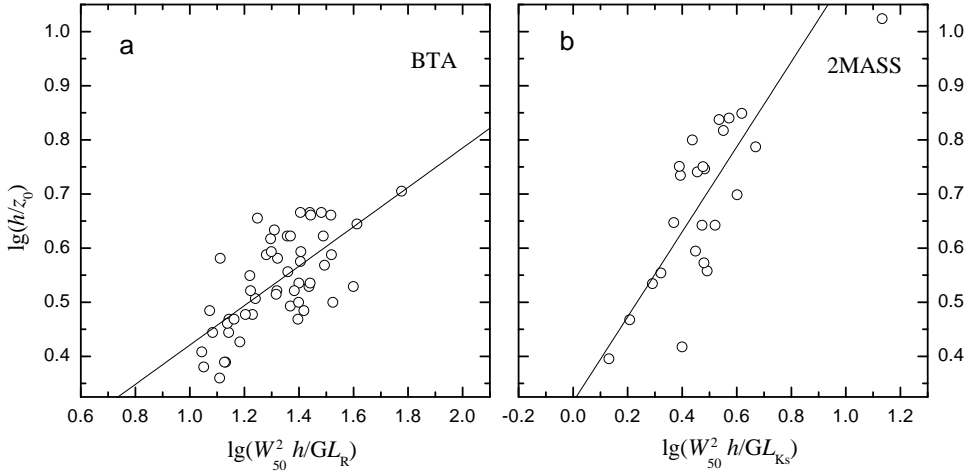} }
\caption{Relation between the photometrically determined radial
to vertical disk scale length ratio and the quantity $W_{50}^2h
/ GL_{R,K}$, which estimates the total mass-to-luminosity ratio
$M/L$ within $R=4h$: a) for the R-band (the BTA sample); b) for
the $K_s$-band (the 2MASS sample).}
\end{figure}

One can see that, although the dispersion is rather high, the
stellar disks indeed tend to be thicker when $M/L$ is lower. It
is better seen for the 2MASS sample, evidently due to the more
reliable photometric models and low internal absorption in the
infrared band. The existence of this dependence is what we
expect if the bending instability is the major factor determining
the disk thickness.

\begin{figure}
\centerline{ \includegraphics[width=10cm]{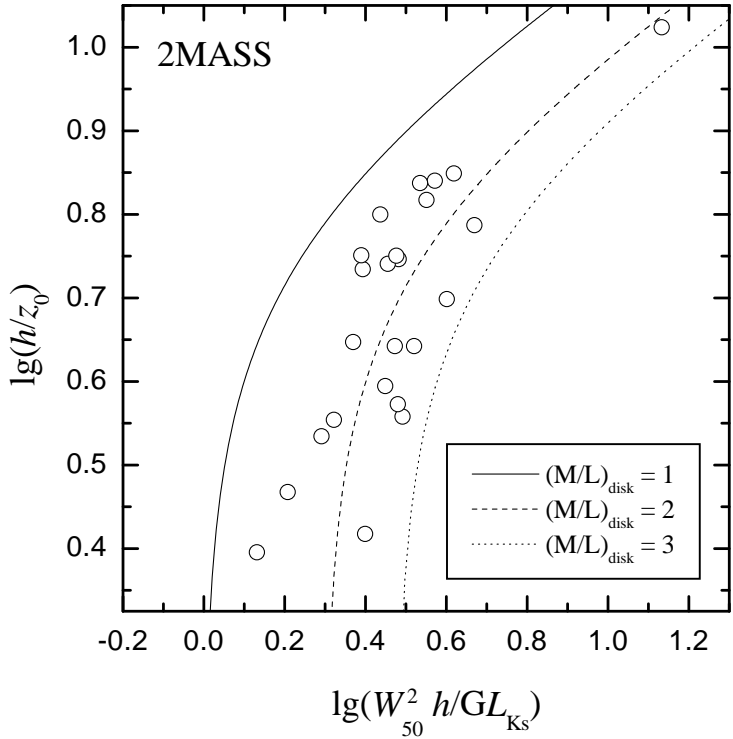} } 
\caption{The same diagram as in Fig. 3b with superposed curves, 
obtained from the numerical simulations (the curve in Fig~2, 
recalculated for three adopted values of disk mass-to- 
luminosity ratios: $(M/L_{Ks})_{disk} $ = 1, 2 and 3 solar units.}
\end{figure}

Fig. 4 is the same as Fig. 3b with three curves superposed. These
curves are what we expect from the N-body numerical simulations
for the initially unstable collisionless galactic disks. The
prototype of the curves is the model curve shown in Fig.~2 above.
It was recalculated to jump from  $M_s/M_d$ to the total ratio
$M/L_{Ks} = W_H^2h/GL_{Ks}$ using the values $(M/L_{Ks})_{disk}
=$ 1, 2 and 3 solar units for a stellar disk.

One can see that the sample galaxies are actually situated in the
domain between $(M/L_{Ks})_{disk}$ =~ 1 and 3 which is quite
reasonable for old stellar systems. It indicates that the
numerical models of marginally stable disks agree well with
observations. In turn, it means that for most of the galaxies the
mechanisms of additional disk heating along Z-axis (other than
the bending instability) may have a little effect on the vertical
disk structure. In this case the approximately constant disk
thickness along the radius is just a result of two opposite
tendencies: the radial decreasing of both the surface disk
density and the vertical velocity dispersion. The influence of
these factors on the vertical scale length cancels each other
almost exactly within the extended range of radial distances.

\begin{figure}
\centerline{ \includegraphics[width=13cm]{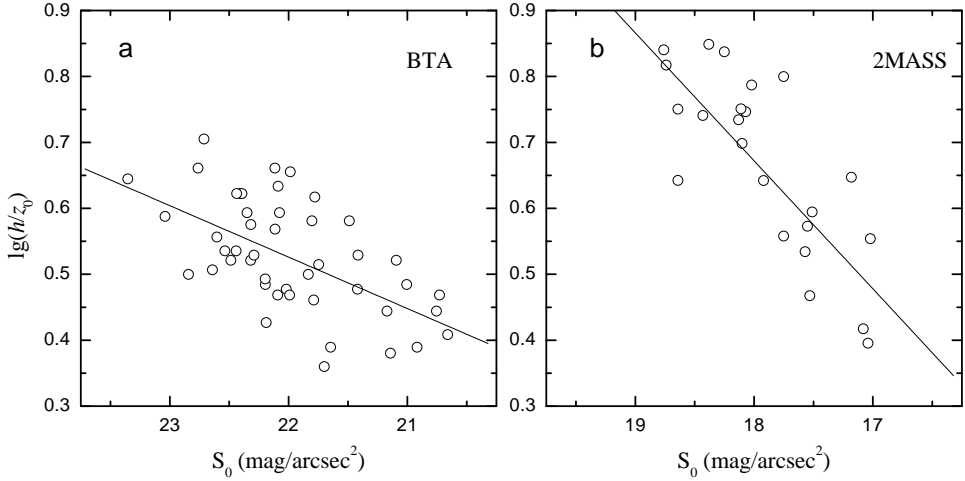} } 
\caption{
Relationship between the radial to vertical disk scalelengths
ratio and the de-projected disk central surface brightness $S_0$
(in magnitudes) for a) the BTA sample and b) the 2MASS sample.}
\end{figure}

Fig. 5 compares the relative disk thickness  and the face-on
central surface brightness (the latter was reduced to face-on
orientation using the model parameters found from the $R$- and the
$K_s$-band surface brightness distributions). The correlation
between these parameters is even more conspicuous than that
between $Z_0/h$ and $M/L$ in Fig~3a,b. Note that the existence of
such a relation is nothing unexpected if to take into account
that the "normal" and the low surface brightness galaxies were
already shown to exhibit the close relation between $S_0$ and the
integrated $M/L$, which characterizes the dark halo mass fraction
(see discussion in MacGaugh \& de Block \cite{mcga98}). The lower
is the central surface brightness (and, consequently, the surface
density), the higher is its flatness and hence the dark halo mass
fraction in a galaxy (the latter dependence between the brightness
and  dark halo mass was also discussed by MacGaugh \& de Block
\cite{mcga98}).

\section{Where do the old stellar disks end?}

The connection between the radial extent of old stellar disks and
their kinematic properties is not so evident as in the case of
their thickness. It is well known that many late type galaxies
experience the significant steepening of the photometric profiles at
some radius (truncation radius) $R_{cut}$ (usually between 2.5
and 5 of photometric scalelengths $h$). As in the case of the
disk flattening, the truncation of the disk can be studied most
accurately in edge-on galaxies. The nature of the disk truncation is
not well known. Here we meet the well known paradox - the less we
know about some event, the easier we find its possible
interpretations. Some authors consider the disk truncation as a result of a
gas density treshold, beyond which the thermal instability, being
responsible for the formation of gas cold phase, is absent. The
others tie the observed edge of a disk with condition of the local
gravitational instability of gaseous layer (see the references in
Pohlen \etal \cite{poh00}, Bizyaev \& Zasov \cite{bizy02}). The
scarcity of the data makes it difficult to verify the different
scenarios. Both approaches mentioned above may explain more or
less successfully the radius where the observed present day star
formation rate drops down in nearby galaxies. Nevertheless, this
radius do not coincide with the truncation radius of an old
stellar disk.

It is rather difficult to verify the role of different mechanisms
regulating the star formation rate in the past when the presently
observed old stellar disk has formed: the velocity dispersion and
the density of a gaseous proto-disk, not even saying about the
ultraviolet ionizing field, should strongly differ from those we
have today. It is possible however to check the fulfilment of the
gravitational stability condition for the proto-disks near the
truncation radius. For this purpose we used the fact that the
critical density of the primeval gaseous disk depends on the same
parameters as the observed thickness of the old stellar disk. In
other words, we admit that the current stellar vertical velocity
dispersion at the truncation radius is equal to (or, in general
case, is not higher than) the velocity dispersion of the parent
gas. Indeed, the critical surface density of a thin gaseous disk
is $\Sigma_{crit} \approx C_{gas}\kappa / Q_T\pi G$ and
$\Sigma_{disk} \approx C_z^2 / \pi G Z_0 \cdot F(\rho_{halo})$,
where $Q_T$ is the Toomre stability parameter, and
$F(\rho_{halo})$ is the coefficient taking into account the
influence of a halo on the disk thickness. To estimate
$\Sigma_{crit}$, we used Polyachenko \etal (\cite{pol97})
criterium which gives $Q_T \approx \sqrt{3}$ for a flat rotation
curve.

We analysed available data for 16 galaxies with the truncation
radii and vertical scalelength values taken from van der Kruit \&
Searle (\cite{vdc81a}, \cite{vdc81b}, \cite{vdc81c} (NGC~891,~
4013,~ 4217, ~4244,~ 4565,~5907,~ 5907), and from Pohlen \etal
\cite{poh00} (NGC~2424, ESO187-008, 269-015, 319-026, 321-010,
446-018, 446-044, 528-017 and 581-006), and whose rotation curves
were found in the literature. The rotation curves were used to
evaluate $\Sigma_{disk}$, $F(\rho_{halo})$ and $\kappa$ at $R =
R_{cut}$ for the galaxies chosen. It is well known that the
rotation curve decomposing, which we used to find
$\Sigma_{disk}$, is ambiguous. Indeed, a comparison of the best
fit model with the maximum disk and the minimum disk models (the
latter is chosen as the model where ($M/L_{Ks})_{disk}$ of the disk
is not less than 0.5) shows that the typical uncertainties for the
disk mass estimates are about 30\%.

\begin{figure}
\centerline{ \includegraphics[height=10cm]{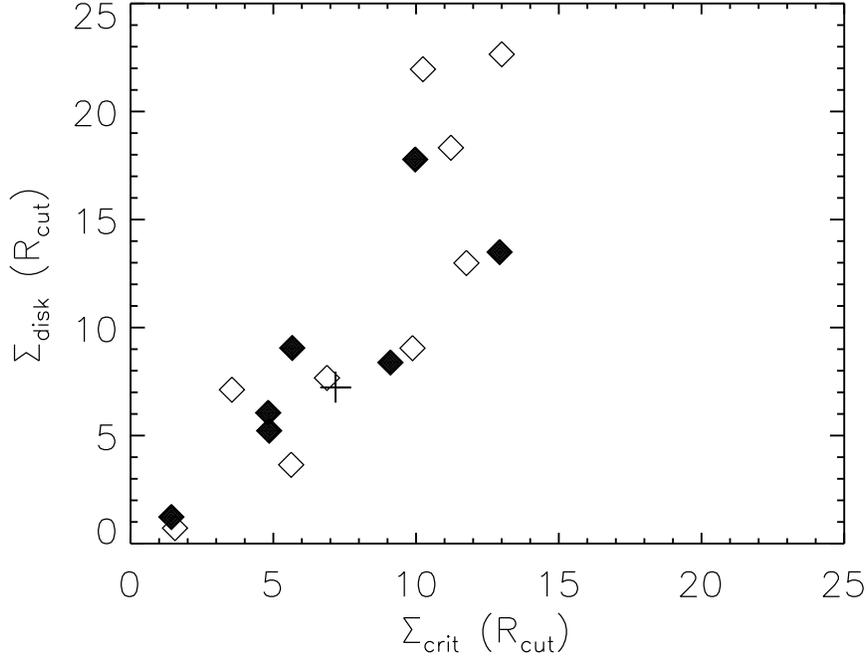} }
\caption{(a) $\Sigma_{disk}$ vs $\Sigma_{crit}$ relation at the
truncation radii $R_{cut}$ for the galaxies taken from the two different
sources (see the text); our Galaxy is marked by the cross.}
\end{figure}
The calculations of $\Sigma_{disk}$ and $ \Sigma_{crit}$ showed
that their ratio decreases slowly toward the periphery for all
galaxies considered, and is close to unit (within a factor of
about 1.5) at $R = R_{cut}$. Their values at $R = R_{cut}$ are
shown in Fig~6. The filled symbols are related to the galaxies
studied by van der Kruit \&Searle (\cite{vdc81a}, \cite{vdc81b},
\cite{vdc81c}), whereas the open ones mark the more distant
galaxies taken from Pohlen \etal~ (\cite{poh00}). The cross marks
the position of our Galaxy.

Curiously, the mean velocity dispersion of stars at the
truncation radius for the galaxies we considered appears to be
close to the turbulent gas velocity in the present interstellar
medium for star forming galaxies: 10- 15 km/s, although it varies
from one galaxy to another.

A comparison of the values $\Sigma_{disk}$ and $ \Sigma_{crit}$
enables us to conclude that the observational data well agree
with the assumption that the observed disk truncation in late
type galaxies considered here is connected with the disk
kinematics. It supports an idea that the sharp decrease of the
surface brightness at $R~>~R_{cut}$  might be a result of rather
inefficient star formation beyond this radius due to the
gravitational stability of outer parts of a primeval gaseous
disk at the time of intense formation of the first generation
stars.


\section{Conclusions}

The thickness of stellar disks and their radial extension in most
of galaxies are rather conservative parameters reflecting the
conditions of the initial formation of galactic disks rather than
their long-lasting evolution. In this case the observed
dependence of the disk thickness on the relative mass of the dark
halo may be naturally explained. The low surface brigtness
galaxies are to be among those which possesses both the thinniest
disks and the highest dark-to-luminous mass ratioes. Concerning
the galactic truncation radii $R_{cut}$, although we cannot prove
their direct relation to the local Jeans instability of the
primeval gas disk at $R< R_{cut}$, however this assumption
demonstrates a good compatibility with the observed properties of
the edge-on galaxies.

This work was supported by the Russian Foundation of Basic Research
(grant 01-02-17597).

\end{document}